\documentclass[aps,prl,superscriptaddress,floatfix,nofootinbib,twocolumn,longbibliography]{revtex4-2}
\usepackage{graphicx}  
\usepackage{titlesec}
\usepackage{subfigure}
\usepackage{dcolumn}   
\usepackage{bm}        
\usepackage{amssymb}   
\usepackage{tabularx}
\usepackage{tablefootnote}
\usepackage{threeparttable}
\usepackage{caption,booktabs}
\usepackage{xcolor}
\usepackage{multirow}
\usepackage[T1]{fontenc}
\usepackage{lipsum,booktabs}
\usepackage[colorlinks,linkcolor=blue,citecolor=blue,anchorcolor=blue]{hyperref}

\begin{document}
\title{
Exploring strong electronic correlations in the breathing kagome metal Fe$_3$Sn}
\author{Shivalika Sharma}
\email{shivalika.sharma@umk.pl}
\affiliation{Institute of Physics, Nicolaus Copernicus University, 87-100 Toru\'n, Poland}
\author{Liviu Chioncel}
\affiliation{Theoretical Physics III, Center for Electronic Correlations and Magnetism, Institute of Physics, University of	Augsburg, 86135 Augsburg, Germany}
\affiliation{Augsburg Center for Innovative Technologies, University of Augsburg, 86135 Augsburg, Germany}

\author{Igor {Di Marco}}
\email{igor.dimarco@physics.uu.se}
\email{igor.dimarco@umk.pl}
\affiliation{Institute of Physics, Nicolaus Copernicus University, 87-100 Toru\'n, Poland}\affiliation{Department of Physics and Astronomy, Uppsala University, Uppsala 751 20, Sweden}
\date{\today}

\begin{abstract}
 Kagome metals have emerged as pivotal materials in condensed matter physics due to their unique geometric arrangement and intriguing electronic properties. Understanding the origin of magnetism in these materials, particularly in iron rich Fe-Sn binary compounds like
Fe$_3$Sn, holds a significant importance, as they represent potential candidates for permanent magnets with a high Curie temperature and a strong magnetic anisotropy. In the present study, we employ density-functional theory and dynamical mean-field theory to analyze the electronic structure and magnetic properties of Fe$_3$Sn.
Our investigation reveals the presence of several nearly-flat bands and Weyl nodes at low excitation energies. The inclusion of local correlation effects is shown to push these features even closer to the Fermi energy, which may be important for their manipulation via external stimuli. Regarding magnetism, the Hubbard-like interaction leads to an increase of orbital polarization at the expenses of a minor reduction of the spin moment. The magnetic anisotropy energy exhibits a strong dependence on the particular choice of the Coulomb interaction parameters.
Additionally, our detailed analysis of the interatomic exchange interactions indicates a significant contribution from the antisymmetric exchange, i.e. the Dzyaloshinskii-Moriya interaction, which showcases the existence of magnetic chirality in the system.
Overall, our investigation highlights a strong interplay between the flat bands near the Fermi level, the local Coulomb interaction and the triangular geometry of the lattice, which  plays a crucial role in driving the magnetic properties of this material.
\end{abstract}
\maketitle

\section{Introduction}\label{sec:intro}
The exploration of kagome lattices has garnered significant interest in condensed matter physics due to their unique geometric configuration that leads to exotic physical phenomena~\cite{yin2022topological,kagome24review}. These lattices are characterized by a pattern of interlaced triangles that create a complex network of hexagons and have emerged as a playground for studying the interplay of electronic topology, strong correlations, and magnetism~\cite{yin2022topological,kagome24review,PhysRevB-AHC}. Kagome materials like AV$_3$Sb$_5$ (A = K, Cs, Rb)~\cite{zhao2021cascade,kagome-review,PRL-Kagome,PhysRevB.108.174413} and Co$_3$Sn$_2$S$_2$~\cite{science.aav2873,liu2018giant,PhysRevB-Co3Sn2S2} were found to host flat bands and Weyl nodes near the Fermi level. Similar features were also reported for the binary metals T$_x$Sn$_y$ (T = Fe, Mn, Co; x:y = 1:1, 3:2, 3:1)~\cite{kuroda2017evidence,ye2018massive,PhysRevLett-flatband,kang2020dirac,chen2022large}.
These characteristics make kagome materials an intriguing platform to witness many exotic phenomena such as topological Hall and Nernst effects, unconventional superconductivity, ferromagnetism, magnetic spin chirality, magnetic skyrmions and more~\cite{yin2022topological,kagome24review}.

The physics of the kagome lattice is further enriched by the possibility of departing from the ideal geometry, allowing for a trimerization that breaks the inversion symmetry. This unique geometry of unequal corner-sharing triangles, known as the breathing kagome lattice, is fascinating due to its potential to host higher-order topological insulators and to exhibit unique magnetic properties~\cite{PhysRevB-breathing,ezawa-PRL}.
Fe$_3$Sn$_2$ provides a material realization of the breathing kagome lattice, with stanene layers sandwiched between Fe$_3$Sn bilayers, thus creating a quasi two-dimensional system. Following reports of a large anomalous Hall effect, a giant magnetic tunability and the presence of skyrmions bubbles, further research on Fe$_3$Sn$_2$ suggested the presence of flat bands, massive Dirac fermions, and a very large Berry curvature in the kagome plane~\cite{ye2018massive}. More detailed experiments via angle-resolved photoemission spectroscopy (ARPES) demonstrated that these Dirac fermions have a dominant contribution from the surface states and Fe$_3$Sn$_2$ behaves instead as a magnetic Weyl semi-metal~\cite{PhysRevB.101.161114}.
Ferromagnetism, reported up to a critical temperature of 670~K, was connected to the presence of flat bands that extend over large sections of the Brillouin zone~\cite{PhysRevLett-flatband}.
The flat bands owe their existence to destructive wave interference - a direct consequence of the unique kagome lattice structure. This interference leads to a quenching of the kinetic energy, making the effect of the electron-electron interaction particularly significant~\cite{Tasaki-flatband}.
For Fe$_3$Sn$_2$, the exchange mechanisms arising from flat bands have been interpreted~\cite{PhysRevLett-flatband} as a coexistence of Mielke's ferromagnetism~\cite{Mielke_1991-flat} and Nakaoka-type ferromagnetism~\cite{Flatband-Nagaoka}.

Flat bands arise also in other Fe-Sn binary systems. FeSn is a metastable material with a simpler stacking, composed by isolated (ideal) kagome planes~\cite{kang2020dirac}. Termination-dependent ARPES measurements detected two-dimensional Dirac cones and flat bands that extend over the full Brillouin zone~\cite{kang2020dirac}.
Conversely, three-dimensional features emerge in the breathing kagome material Fe$_3$Sn. This compound was the first extensively studied system among Fe-Sn binaries, due to a stable ferromagnetic order with a Curie temperature of 705~K and a large magnetic anisotropy energy (MAE)~\cite{sale-et-al,FAYYAZI2019126,Liviu_APL}.
Although these characteristics would make Fe$_3$Sn an ideal candidate as a component for rare-earth free permanent magnets, the easy axis does not lay along the $c$ crystallographic axis, but inside the kagome plane~\cite{Fe3Sn-RSPt}. This {\textit{de facto}} easy plane is not suitable for magnetic applications. Alloying with Sb, As, and Te and nanostructuring have been explored as possible pathways to rotate the easy axis to be perpendicular to the kagome plane, but this approach has had limited success, due to the high sensitivity of the MAE to compositional parameters and electron or hole doping~\cite{Fe3Sn-RSPt}.
More recent works have pointed out that Fe$_3$Sn hosts a large anomalous Hall effect~\cite{PhysRevB-AHC,low2023,prodanPRB2024} and a large anomalous Nernst effect~\cite{kurosawa2024}.
These topological features can be connected to the presence of a nodal plane composed by two nearly flat bands closely facing each other in the two dimensional plane~\cite{chen2022large,kurosawa2024}.
Since the stacking of Fe$_3$Sn forms a genuine three-dimensional network rather than isolated planes, a comprehensive understanding of its electronic structure from first principles becomes crucial. This also includes the analysis of strong electronic correlations, whose importance for kagome-based materials has been recently emphasized by a theoretical study on Sc$_3$Mn$_3$Al$_7$Si$_5$~\cite{Mn-Kagome}, as well as by high-resolution laser-based micro-focused ARPES measurements on Fe$_3$Sn$_2$~\cite{ekahana2024nature}.

In the present study, we have performed an extensive analysis of the electronic and magnetic properties of bulk Fe$_3$Sn, based on a combination of density-functional theory (DFT)~\cite{martin_book} and dynamical mean-field theory (DMFT)~\cite{DMFT}. This DFT+DMFT approach~\cite{kotliar06rmp78,held07ap56} is employed to understand how strong electronic correlations affect the behavior of the flat bands in different parts of the Brillouin zone. The calculation of the interatomic exchange parameters between local moments at the Fe sites makes it possible to connect these effects to the strength of the magnetic coupling. The paper is organized as follows. After this Introduction, the Methodology of our work is presented in Section~\ref{sec:methods}. The results of our research are then illustrated in Section~\ref{sec:magnetism} and Section~\ref{sec:estructure}, separately for the Magnetic properties and the Details of the electronic structure. Finally, the Conclusions of our work are presented in Section~\ref{sec:conclusions}.

\section{Methodology}\label{sec:methods}
\begin{figure*}
\centering
\includegraphics[scale=0.4]{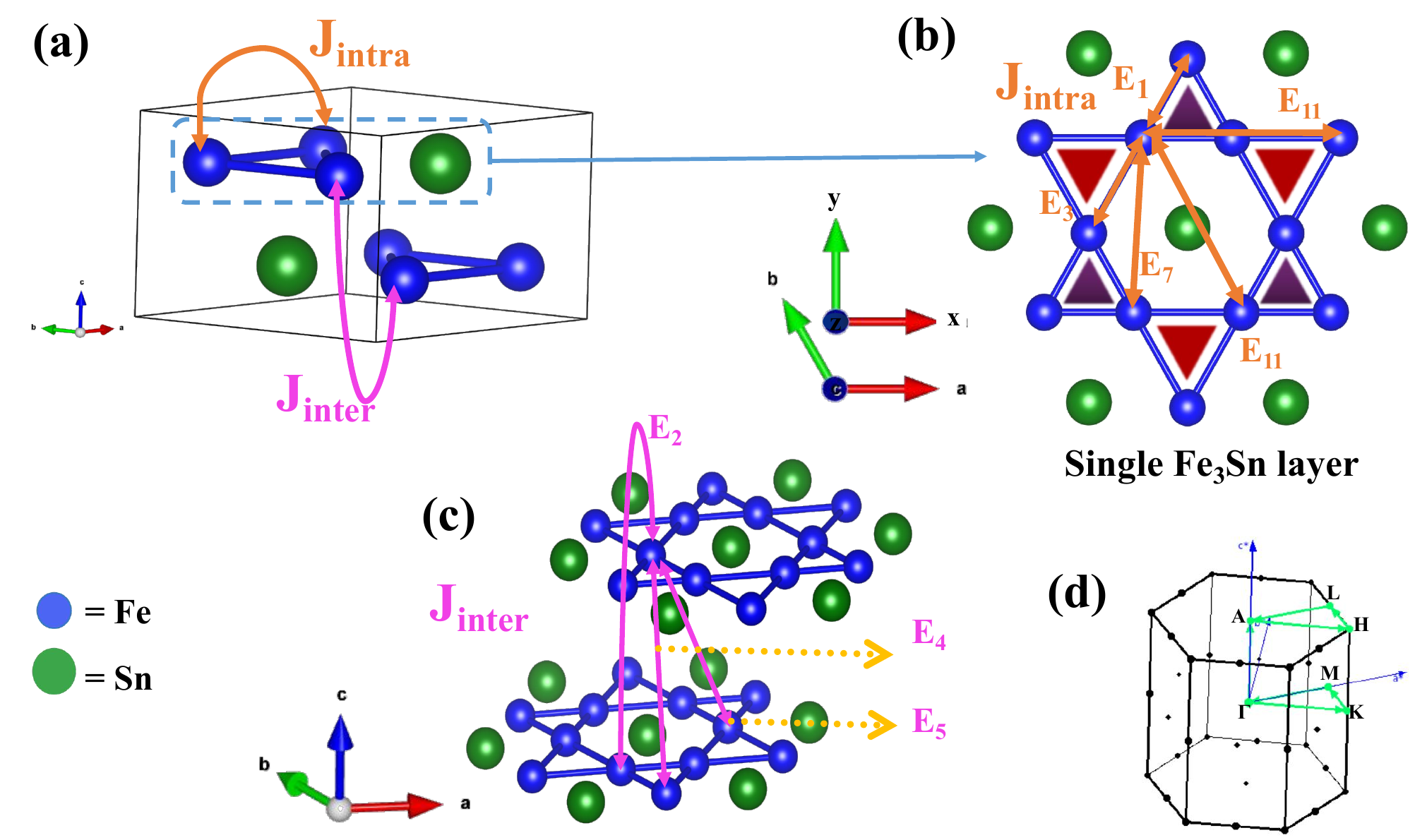}
\caption{\label{fig:first}(a) Illustration of the Fe$_3$Sn unit cell, highlighting interatomic exchange paths between two Fe atoms. (b) A single kagome layer (Fe$_3$Sn), consisting of two differently sized equilateral triangles (indicated in dark red and violet), with selected intralayer exchange paths. (c) Stacking of two kagome layers in the P6$_3$/mmc structure, with selected interlayer exchange paths. (d) The Brillouin zone of the hexagonal kagome lattice showing paths connecting high symmetry points.}
\end{figure*}
We have considered the bulk crystal structure of Fe$_3$Sn, consisting of layers of Fe and Sn in the A-B stacking, as shown in Fig.~\ref{fig:first}(a). It has a space group of P6$_3$/mmc, with Fe atoms forming the kagome lattice and Sn atoms residing in the centre of the hexagon, on a different plane along the $c$ crystallographic axis. The structural data used for the calculations were taken from experiments on single-crystal samples~\cite{Liviu_APL}. The corresponding lattice parameters are $a=b=5.4604$~{\AA} and $c= 4.3458$~{\AA}, while the breathing distortion gives a ratio of approximately 0.86 between the sides of the smaller and larger triangles.
These triangles of different sizes, emphasized by red and violet shades in Fig.~\ref{fig:first}(b), are the fundamental components of the breathing kagome lattice, corresponding to a single Fe$_3$Sn layer.
DFT and DFT+DMFT calculations were performed by using the full potential linear muffin-tin orbital method (FP-LMTO) as implemented in the Relativistic Spin-Polarized Toolkit (RSPt) code~\cite{RSPt,GRANAS2012295}. The generalized gradient approximation (GGA) was used for the exchange-correlation functional, based on the Perdew and Wang functional with Perdew-Burke-Ernzerhof (PBE) corrections~\cite{perdew_PhysRevLett.77.3865_1996,perdew_PhysRevLett.78.1396_1997,pw92}.
For the DFT+DMFT calculations, the effective impurity problem has been solved through the spin-polarized T-matrix fluctuation-exchange (SPTF) solver~\cite{katsnelson02epjb}, in its spin- and orbital-rotationally invariant formulation~\cite{prb-fe-Ni-DFT-DMFT,PhysRevB-sptf}.
SPTF is a perturbative solver that works well for magnetic systems with moderate correlations, but is not adequate for strong many-body effects such as those associated to the formation of fluctuating local moments in the paramagnetic phase~\cite{dimarco09epjb72473}, the emergence of Kondo physics~\cite{litsarev12prb86_115116} or the Mott metal-to-insulator transition~\cite{drchal05}.
SPTF has proved to be very successful in describing electronic and magnetic properties of transition metal elements and compounds, including itinerant ferromagnets~\cite{PhysRevB-sptf,braun06prl,dimarco09prb79}, half-metallic ferromagnets~\cite{katsnelson08rmp}, two-dimensional magnets~\cite{JIJ-MAE-RSPt}, high-entropy alloys~\cite{Redka2024}, as well as kagome systems with complex magnetic textures~\cite{PRL-Kagome,PhysRevB.108.174413}.
Thus, we expect SPTF to be perfectly adequate to describe Fe$_3$Sn as well. Like in our previous studies, the dynamical self-energy is obtained as a function of the renormalized Hartree-Fock Green's function~\cite{PRL-Kagome,PhysRevB.108.174413}.
Moreover, when the effective impurity problem is solved in the Hartree-Fock approximation, we recover the DFT+U approach~\cite{Vladimir_1997,kotliar06rmp78}.
The double counting term for DFT+U is chosen as the around mean field (AMF) correction~\cite{Vladimir_1997,kotliar06rmp78}. For DFT+DMFT, the orbital average of the static part of the self-energy is used~\cite{dimarco09prb79}, which reduces to the AMF correction if only Hartree-Fock terms are included.
The basis of the local orbitals is constructed by considering only the muffin-tin heads, as described in Ref.~\cite{PhysRevB-sptf}.
The Coulomb interaction term, to be solved in DFT+U and DFT+DMFT, is applied to the Fe-$3d$ states only, using the aforementioned basis.
Regarding the Coulomb interaction parameters, no estimates can be found in literature specifically for Fe$_3$Sn. Since bandwidth and band filling are not too different than those of bulk bcc Fe, we decided to employ the Coulomb interaction parameters used in the past studies for this system~\cite{PhysRevLett.87.067205,prb-fe-Ni-DFT-DMFT,PhysRevLett.106.106405}. Hence, the Hubbard $U$ and the Hund exchange $J$ to be used in DFT+DMFT have been set to 2.3 eV and 0.9 eV. These values are also compatible with recent calculations made for FeSn by means of the constrained random phase  approximation (cRPA)~\cite{PhysRevResearch.5.L012008}. While the Hubbard U predicted there is slightly higher (3~eV), one should keep in mind that the hybridization in Fe$_3$Sn is larger, due to the presence of a stronger out-of-plane coupling.
For DFT+U calculations, the aforementioned values have been decreased to 1.5 eV and 0.8 eV, to account for the fact that in the Hartree-Fock solution there is no dynamical screening~\cite{dimarco18prb97}.
This reduction is somehow arbitrary, but it is necessary to have a meaningful comparison with DFT+DMFT. Nevertheless, this arbitrariness implies that the comparison between DFT+U and DFT+DMFT can only be made at a qualitative level.
Other choices for U and J have been explored, as commented throughout the text.
The spectral functions are calculated via Green's function theory at a distance of 13~meV from the real energy axis~\cite{prb-fe-Ni-DFT-DMFT}, unless stated otherwise. The analytical continuation was performed directly on the self-energy function by means of the revised Pad\'e approximant method~\cite{PhysRevB.93.075104}.

Converged electronic structure calculations from DFT, DFT+U, and DFT+DMFT are then utilized to extract interatomic exchange parameters $J_{ij}$ by mapping the magnetic excitations onto to an effective Heisenberg Hamiltonian~\cite{Igor_PhysRevB_2015}:
\begin{equation}
 {H} = - \sum_{i\neq j} J_{ij}\vec{e_{i}}\cdot\vec{e_{j}}
 \label{eqn1}
\end{equation}
 Here $(i,j)$ are the indices for atomic sites hosting the magnetic moments, while $\vec{e_i}$ and $\vec{e_j}$ are unit vectors along the spin direction at sites $i$ and $j$, respectively. The $J_{ij}$ are calculated as the response to an infinitesimal rotation of two spins at sites $i$ and $j$, by means of a generalized magnetic force theorem~\cite{Igor_PhysRevB_2015,lichtenstein87jmmm67}. In presence of spin-orbit coupling (SOC), the interatomic exchange becomes anisotropic and Eqn.~\ref{eqn1} can be rewritten in a tensorial form, which can be recast as~\cite{KvashninPhysRevB_2020}:
\begin{eqnarray}
  {H} = - \sum_{i\neq j}   J_{ij}\vec{e_{i}}\cdot\vec{e_{j}} -  \sum_{i\neq j}  \mathbf{e}_{i}^T \mathbb{J}_{ij}^S \mathbf{e}_{j} -  \sum_{i\neq j}  \vec{D}_{ij}\cdot(\vec{e_{i}}\times \vec{e_{j}})
  \label{eqn2}
\end{eqnarray}
Here, the first term is the isotropic exchange interaction corresponding to Eqn.~\ref{eqn1}, the second term is the symmetric anisotropic exchange and $\mathbb{J}^S_{ij}$ is the symmetric anisotropic exchange tensor. The third term is the antisymmetric anisotropic exchange, also known as the Dzyaloshinskii–Moriya (DM) interaction~\cite{Moriya_PhysRev_1960,DZYALOSHINSKY1958241}. The scalar triple product is written in terms of the DM vector $\vec D_{ij}$, whose explicit expression can be found in Ref.~\onlinecite{PhysRevB.68.104436}. Usually the strength of the DM interaction is discussed with respect to the magnitude of the DM vector:
\begin{equation}
    |\vec D_{ij}|=\sqrt{{(D^x_{ij})}^2+{(D^y_{ij})}^2+{(D^z_{ij})}^2}
\end{equation}
 where $\vec{D^x}$, $\vec{D^y}$ and $\vec{D^z}$ are the directional components of $\vec D_{ij}$ along the Cartesian axes. In our reference frame, the crystallographic directions are arranged to have $x$ and $z$ parallel to $a$ and $c$, respectively, as also shown in Fig.~\ref{fig:first}(b).
 Finally, the sampling of the Brillouin zone was performed by a dense Monkhorst-Pack grid of $24 \times 24 \times 24$ $\mathbf{k}$-points, in order the ensure the convergence of the interatomic magnetic exchange coupling. All calculations were performed for a temperature of 200~K.

\section{Magnetic properties}\label{sec:magnetism}
Previous computational works have demonstrated that the ground state of Fe$_3$Sn is ferromagnetic with an easy plane~\cite{sale-et-al,Fe3Sn-RSPt}. Our DFT calculations confirm these results and predict a MAE of 0.64 meV per formula unit between the easy axis (100) and the hard axis (001).
In the easy plane, the calculated energy differences amount to less than 2 $\mu$eV. Such small differences are expected due to the hexagonal symmetry, which makes spin orientations within the a-b plane energetically degenerate.
The MAE resulting from our calculations, which in different units amounts to 1.8 MJ/m$^3$, is in perfect agreement with the value reported in the recent work by Belbase \textit{et al.}, based on the full-potential local-orbital (FPLO) method~\cite{PhysRevB-AHC}.
Other theoretical studies~\cite{sale-et-al,FAYYAZI2019126,PhysRevB-AHC,Liviu_APL} predicted values that range from 1.3 MJ/m$^3$ to 1.6 MJ/m$^3$, which is slightly larger than the discrepancy that one may expect nowadays from electronic structure calculations~\cite{lejaeghere16}.
This discrepancy has been already pointed out as a consequence of the high sensitivity of the MAE to the c/a ratio, which may greatly amplify minor structural differences, as e.g. due to structural optimization or experimental input~\cite{Fe3Sn-RSPt,PhysRevB-AHC}. The most recent experimental results on single crystals~\cite{low2023,Liviu_APL} estimate the MAE to be around 1.0 MJ/m$^3$ at room temperature and then to increase up to 1.3 MJ/m$^3$ at 2~K. These values are from 30\% to 40\% smaller than the number we obtained from electronic structure calculations.
A similar discrepancy characterizes the saturation magnetization, whose measurements give values in the range of 6.6-6.8~$\mu_{\text{B}}$ per formula unit~\cite{FAYYAZI2019126,low2023,Liviu_APL}. Our DFT calculations give a slightly larger value amounting to 7.1~$\mu_{\text{B}}$ per formula unit, which is in perfect agreement with previous theoretical studies~\cite{sale-et-al,Fe3Sn-RSPt,FAYYAZI2019126,low2023,Liviu_APL,PhysRevB-AHC}. Site-projected magnetic moments are reported in Table~\ref{tab:gs_magmoments}, including also spin and orbital decompositions.
These data show that the Fe and Sn spin moments are anti-parallel to each other and that the contribution of the orbital magnetism is rather small, in agreement with itinerant metallic character of this material. These values are in excellent agreement with those obtained in FPLO calculations, especially if we take into account the usage of different sets of local orbitals for the site projections~\cite{FAYYAZI2019126}.

\begin{table}[t]
\begin{tabular}{lccccc}
\hline
     & \: $\mu_{\text{tot}}$\: &\: $\mu^s_{Fe}$ \: & \:  $\mu^o_{Fe}$ \:  & \:  $\mu^s_{Sn}$ \:  & \:  $\mu^o_{Sn}$ \:   \\
\hline
DFT               &  7.14 & 2.45 & 0.07 & -0.11 & 0.00  \\
DFT+U             &  6.93 & 2.34 & 0.12 & -0.13 & 0.00  \\
DFT+DMFT (def)    &  6.82 & 2.31 & 0.10 & -0.12 & 0.00  \\
DFT+DMFT (red)    &  6.93 & 2.35 & 0.09 & -0.12 & 0.00  \\
\hline
\end{tabular}
\caption{Total magnetic moments per formula unit and site-projected spin and orbital moments for bulk Fe$_3$Sn, as computed in DFT, DFT+U and DFT+DMFT, using the default Coulomb interaction parameters, as well as reduced values $U=2.0$~eV and $J=0.65$~eV. The values are given in $\mu_{B}$.\label{tab:gs_magmoments}}
\end{table}

Next, we focus on how the magnetic properties are affected by the inclusion of an explicit Hubbard interaction term, via DFT+U and DFT+DMFT. Regarding the MAE, the value obtained from DFT+U calculations amounts to 2.4 MJ/m$^3$. This increase with respect to the DFT result is mainly driven by the increase in the orbital polarization of the Fe-$3d$ states, as evident from Table~\ref{tab:gs_magmoments}. These results are consistent with the even larger MAE that was recently obtained~\cite{PhysRevB-AHC} through the inclusion of the orbital-polarization correction~\cite{LNordstrom1992}. The rise of the orbital moment observed in Table~\ref{tab:gs_magmoments} is accompanied by a small decrease of the spin moment, which is a consequence of the very definition of the AMF double counting for DFT+U~\cite{Ylvisaker09}. The balance between spin and orbital magnetism is supposed to be better described when magnetic fluctuations that go beyond the static mean-field approximation are included~\cite{kotliar06rmp78,PhysRevB-sptf}. The DFT+DMFT results reported in Table~\ref{tab:gs_magmoments} show that dynamical correlation effects drive a further decrease of the spin moment but also reduce the orbital moment with respect to the DFT+U values. The MAE calculated with these (default) settings is however extremely large, amounting to 5.9 MJ/m$^3$. Additional calculations reveal that the MAE estimated from DFT+DMFT is very sensitive to small variations of the Coulomb interaction parameters, in particular the Hund's exchange $J$. If the chosen parameters, namely $U=2.3$~eV and $J=0.9$~eV, are decreased to $U=2.0$~eV and $J=0.65$~eV, then the MAE reduces to 1.3 MJ/m$^3$, in agreement with the experiment, while the spin and orbital moments become closer to the original DFT values, see Table~\ref{tab:gs_magmoments}. More comprehensive data on the dependence of the MAE on the Coulomb interaction parameters are presented in the Supplementary Material (SM)~\cite{supplemental}.
Overall, the large variations of the MAE may have a methodological origin, reflecting e.g. the deficiencies of approaches where the Hubbard correction is applied on top of a spin-polarized DFT description~\cite{keshavarz18,Jang18,JIJ-MAE-RSPt}. Alternatively, they may have a physical origin, connecting to the high sensitivity of the MAE to small structural differences, as emphasized above.
The latter interpretation would be in line with the expectation of a strong interplay between electronic correlations, magnetism and topology on the unique kagome geometry~\cite{shen2022thermodynamical,ekahana2024nature,huang2020,PRL-Kagome,Mn-Kagome}.
Recent findings on FeSn, approaching the limit of a two-dimensional kagome lattice, have also emphasized orbital selective physics driven by the Hund's exchange, in analogy to iron-based superconductors~\cite{Ren2024}. Finally, the reduction of the total magnetic moment observed in both DFT+U and DFT+DMFT with respect to DFT is also more consistent with the measured saturation magnetization of 6.6-6.8~$\mu_{\text{B}}$ per formula unit~\cite{FAYYAZI2019126,low2023,Liviu_APL}.

\begin{figure}[t]
    \includegraphics[scale=0.61]{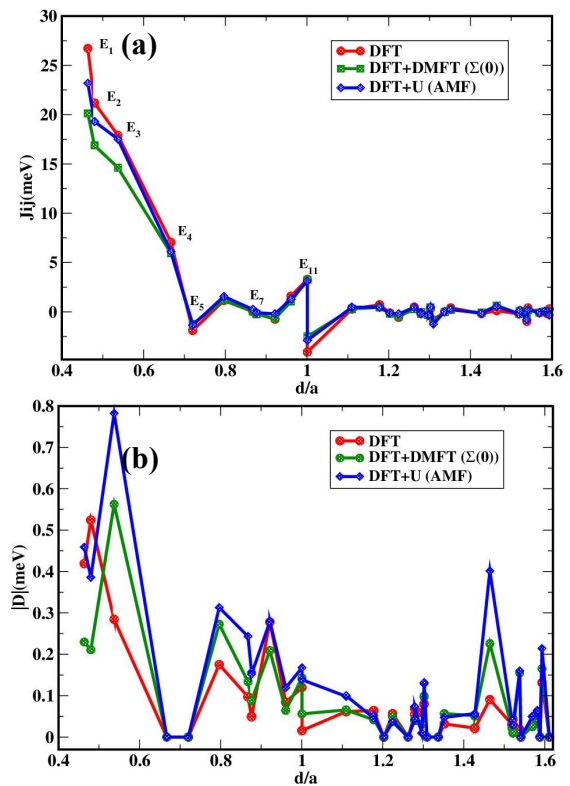}
    \caption{(a) Isotropic averages of the exchange tensor as a function of the distance between Fe atoms. (b) The magnitude of the DM vector as a function of the distance between Fe atoms. Calculations are done with RSPt in DFT, DFT+DMFT and DFT+U with the inclusion of SOC.\label{fig:jijs}}

\end{figure}

Having established the basic properties of the magnetic ground state, we proceed to the analysis of the interatomic exchange coupling, according to Eqn.~\ref{eqn2}.
In the Fe$_3$Sn unit cell shown in Figure~\ref{fig:first}(a), we highlight the difference between the two fundamental groups of exchange interactions, namely the intralayer terms J$_{\text{intra}}$ and the interlayer terms J$_{\text{inter}}$.
The most important exchange paths E$_n$ between two Fe atoms are illustrated in Figure~\ref{fig:first}(b) and Figure~\ref{fig:first}(c), for J$_{\text{intra}}$ and J$_{\text{inter}}$ respectively. The index $n$ labels the shell of nearest neighbors (NNs), ordered with respect to their interatomic distance.
 With this in mind, we can inspect the plots of the isotropic averages and the DM interaction as a function of the interatomic distance, reported in Figure~\ref{fig:jijs}.
In standard DFT, the isotropic exchange interaction for the first four shells of NNs is markedly ferromagnetic, in agreement with what we have discussed above. The exchange coupling for the first NN shell is the largest, and corresponds to the intralayer path E$_1$, connecting two Fe atoms in the smaller triangle of the breathing kagome. The coupling between Fe atoms in the larger triangle, along the path E$_3$, is about 30\% smaller than the first NN value. The second and fourth NNs are located along the interlayer paths E$_2$ and E$_4$, and yet they also exhibit a substantial ferromagnetic coupling, in agreement with the fact that the distance between the kagome planes (2.61 {\AA}) is comparable to the side of the triangles in the breathing kagome (2.52~{\AA} and 2.93~{\AA}).
The exchange interaction of the first four shells of NNs demonstrates quantitatively that magnetism in Fe$_3$Sn arises from a full three-dimensional network, in agreement with the experimental trends of the Fe stannides~\cite{FAYYAZI2019126,kang2020dirac,kurosawa2024}.
The first antiferromagnetic coupling emerges for the fifth shell of NNs, along the interlayer path E$_5$, shown in Figure~\ref{fig:first}(c).
For larger interatomic distances, the system exhibits a long-range oscillatory behavior that is reminiscent of the Ruderman–Kasuya–Kittel–Yosida (RKKY) interaction~\cite{blundell01bookpg1567}. In Figure~S2 of the SM~\cite{supplemental}, we show that, in standard DFT, the interatomic exchange interaction approximately scales as the cube of the interatomic distance $d$. This scaling is precisely what is expected from RKKY coupling in a three-dimensional system~\cite{blundell01bookpg1567}, and offers a definitive support to the three-dimensional nature of the ferromagnetism in Fe$_3$Sn. A more quantitative theory of the RKKY oscillations would require an analysis of the caliper vectors defined in the Fermi surface, as in Ref.~\onlinecite{Sarkar_2022}, but this is beyond the scope of the present article. Interestingly, the role of the RKKY coupling in layered topological semimetals has been recently pointed out, albeit for non-coplanar magnetism involving different sublattices~\cite{RKKY-Kagome}.
Going back to the analysis of Figure~\ref{fig:jijs}(a), competing ferromagnetic and antiferromagnetic interactions are visible for the eleventh shell of NNs, indicating the presence of frustration in the system due to its structural geometry, which leads to two different paths E$_{11}$, as depicted in Figure~\ref{fig:first}(b).

The amplitude of the DM vector as a function of the interatomic distance is shown in Figure~\ref{fig:jijs}(b), while the components D$^x$, D$^y$, and D$^z$ for selected exchange paths are given in Table~\ref{tab:table2}. The largest coupling is interlayer, connecting Fe atoms along the path E$_2$, and amounts to 0.52 meV, mostly arising from its D$^z$ component. Inside the kagome plane, the first and second shells of NNs are characterized by slightly smaller DM vectors, whose major contributions still come from the D$^z$ component. These finite values of the DM interaction indicate the tendency for chirality between two Fe sites with the preferred direction for the DM vector along the z-axis (out-of-plane). The magnitude of the DM interaction is relatively small ($\sim$3-4\%) compared to the isotropic Heisenberg exchange tensor, making the likelihood of spin canting quite low~\cite{CAMLEY2023}. Definitive confirmation of collinear order would require the solution of the effective Heisenberg Hamiltonian defined by Eqn.~\ref{eqn2}, which would be an interesting task for future works involving multi-scale modeling~\cite{book_olle}.
Furthermore, we see that |D| for the forth and fifth NN shells is zero, because of the presence of inversion symmetry, as evidenced by the interlayer exchange paths E$_4$ and E$_5$ depicted in Figure~\ref{fig:first}(c).

\begin{table}[b]
    \centering
    \begin{tabular}{|c|c|c|c|c|}
    \hline
    \multirow{2}{*}{\textbf{{\begin{tabular}[c]{@{}c@{}}Exchange  \\ path\end{tabular}}}}& \multirow{2}{*}{\textbf{{\begin{tabular}[c]{@{}c@{}}DM  \\ vector\end{tabular}}}}& \multirow{2}{*}{\textbf{DFT}}& \multirow{2}{*}{\textbf{DFT+DMFT}}&\multirow{2}{*}{\textbf{DFT+U}} \\
    & & & &\\\hline
     \multirow{3}{*}{\textbf{E$_1$}}&\textbf{D$^x$}&0& 0& 0\\
    &\textbf{D$^y$}&0 & 0&0\\
    &\textbf{D$^z$}&0.41&0.22&0.50\\\hline
    \multirow{3}{*}{\textbf{E$_2$}}&\textbf{D$^x$}&-0.08&-0.05&-0.05\\
    &\textbf{D$^y$}&0.04 & 0.02&0.02\\
    &\textbf{D$^z$}&0.51 &0.19&0.36\\\hline
    \multirow{3}{*}{\textbf{E$_3$}}&\textbf{D$^x$}&0&0 &0\\
    &\textbf{D$^y$}&0 & 0&0\\
    &\textbf{D$^z$}&0.28 &0.56&0.78\\\hline
    \multirow{3}{*}{\textbf{E$_6$}}&\textbf{D$^x$}&-0.16&-0.27 &-0.31\\
    &\textbf{D$^y$}&0 & 0&0\\
    &\textbf{D$^z$}&0& 0&0\\\hline
    \multirow{3}{*}{\textbf{E$_9$}}&\textbf{D$^x$}&0.09&0.09 &0.13\\
    &\textbf{D$^y$}&0.16 & 0.17&0.23\\
    &\textbf{D$^z$}&0.19& 0.05&0.07\\\hline
    \end{tabular}
    \caption{Components of the DM vector along the Cartesian axes as depicted in Figure~\ref{fig:first}, as calculated in DFT, DFT+DMFT and DFT+U by RSPt. All values are given in meV.\label{tab:table2}}
\end{table}

\begin{figure}[t]
    \includegraphics[scale=0.35]{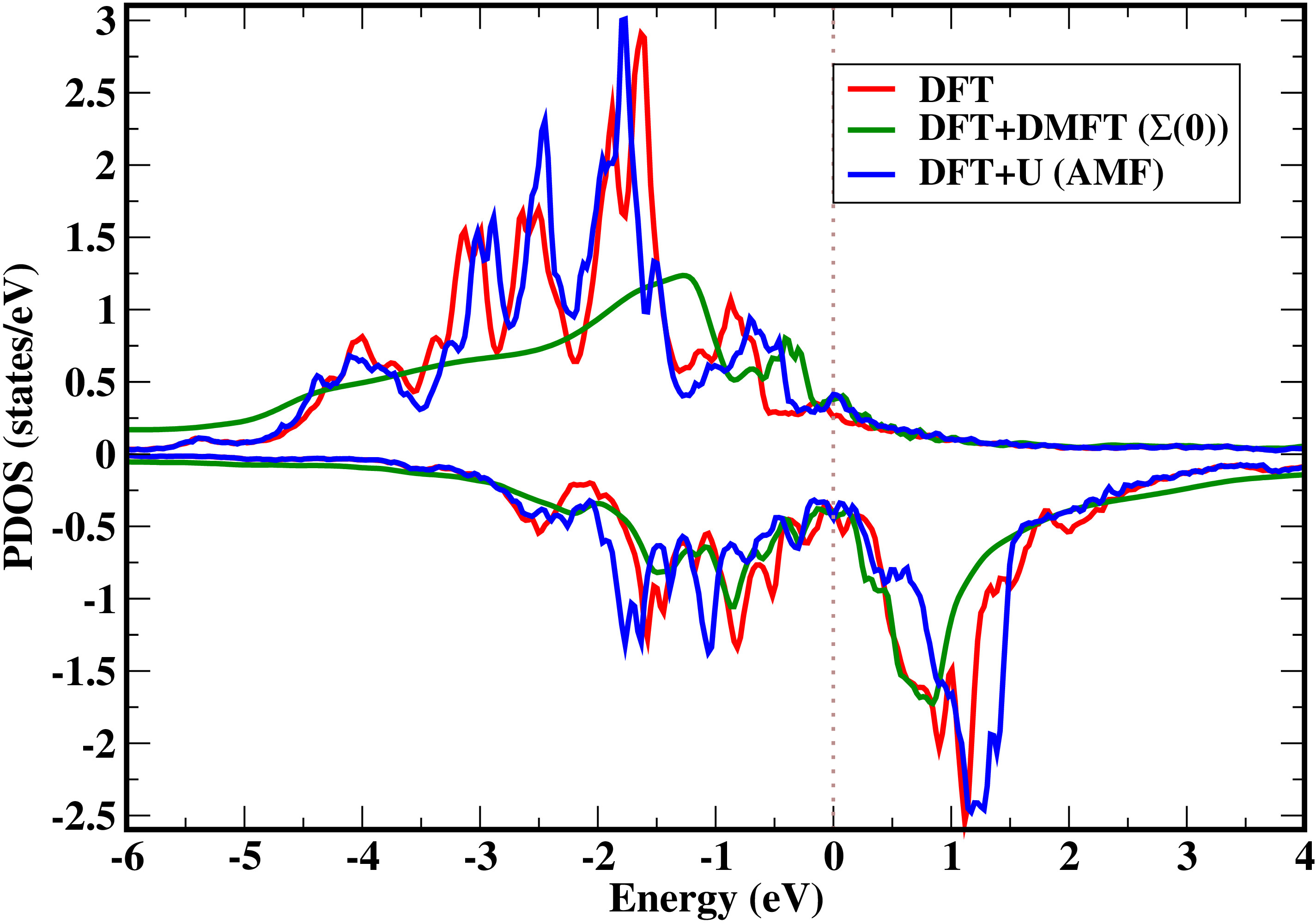}
    \caption{PDOS of Fe-3$d$ states as obtained in DFT, DFT+DMFT and DFT+U via RSPt. The Fermi level is located at zero energy.\label{fig:FePdos}}
\end{figure}

The inclusion of an explicit Hubbard term via DFT+U and DFT+DMFT leads to a small reduction of the isotropic exchange, as shown in Figure~\ref{fig:jijs}(a). This reduction reflects the smaller values of the Fe spin moments reported in Table~\ref{tab:gs_magmoments} through the definition used in Eqn.~\ref{eqn2}.
By inspecting the projected density of states (PDOS) for the Fe-$3d$ states, depicted in Figure~\ref{fig:FePdos}, we observe that the exchange splitting is largest in DFT and smallest in DFT+DMFT, in agreement with the trends of the magnetic moments and interatomic interactions.
The overall reduction of the strength of the magnetic coupling in DFT+DMFT is similar to the one previously observed for bcc Fe~\cite{Chadov2008,Igor_PhysRevB_2015}. Nevertheless, it is surprising that the DFT+U does not cause a stronger localization of the Fe-$3d$ electrons, making the $J_{ij}$'s shorter ranged.
Although a possible reason for the observed behavior may be the usage of smaller Coulomb interaction parameters in DFT+U, additional calculations demonstrate that this is not the leading factor (data not shown).
In fact, the long-range nature is found to be a consequence of the AMF double counting, which is usually used for correlated metals~\cite{Petukhov03}.
To evaluate the impact of the double counting, we performed additional electronic structure calculations in DFT+U in the fully-localized limit (FLL) while keeping the other settings unchanged. As expected, a much stronger electronic localization is observed, resulting in an increase of the exchange splitting (data not shown), a higher local magnetic moment of 2.54 $\mu$B, and an increase of the first NN coupling up to 31 meV. This magnetic moment deviates significantly from the experimentally reported values, which demonstrates that the FLL double counting is not suitable for describing this system, as usual for itinerant ferromagnets~\cite{Petukhov03}.

The inclusion of the on-site Coulomb interaction via DFT+DMFT and DFT+U leads to significant qualitative and quantitative changes in the DM interaction. For both methods, Figure~\ref{fig:jijs}(b) shows a marked increase in the value of |D| for the third shell of NNs, corresponding to the intralayer exchange path E$_3$. The first shell of NNs, along the intralayer exchange path E$_1$, exhibits a more complex trend, as the DM interaction increases in DFT+U but decreases in DFT+DMFT.
The dominant character of the intralayer terms remains D$^z$, compatibly with the symmetry of the kagome lattice.
The DM interaction with the second shell of NNs, along the exchange path E$_2$, decreases for both DFT+U and DFT+DMFT, an effect driven mainly by a change in D$^z$.
Notably, there is also a significant increase in the magnitude of the DM interaction with farther NNs, as e.g. in the sixth and ninth shells, which we attribute to a stronger orbital polarization driven by the inclusion of the on-site Coulomb interaction.
These data, and in particular the decomposition shown in Table~\ref{tab:table2}, suggest deeper changes in the electronic structure than those observed from the PDOS of Figure~\ref{fig:FePdos}. We will investigate these changes in the next section.

For completeness, we also calculated the values of the interatomic exchange interactions for the non-relativistic case, corresponding to Eqn.~\ref{eqn1}. As illustrated in Figure~S1 of the SM~\cite{supplemental}, the isotropic exchange is only marginally affected by the presence of SOC, and the primary consequence of the latter is to induce anisotropic terms as the DM interaction, favoring spin canting and complex magnetic textures~\cite{PhysRevB.108.174413}.

\begin{figure*}
   \includegraphics[width=1\textwidth]{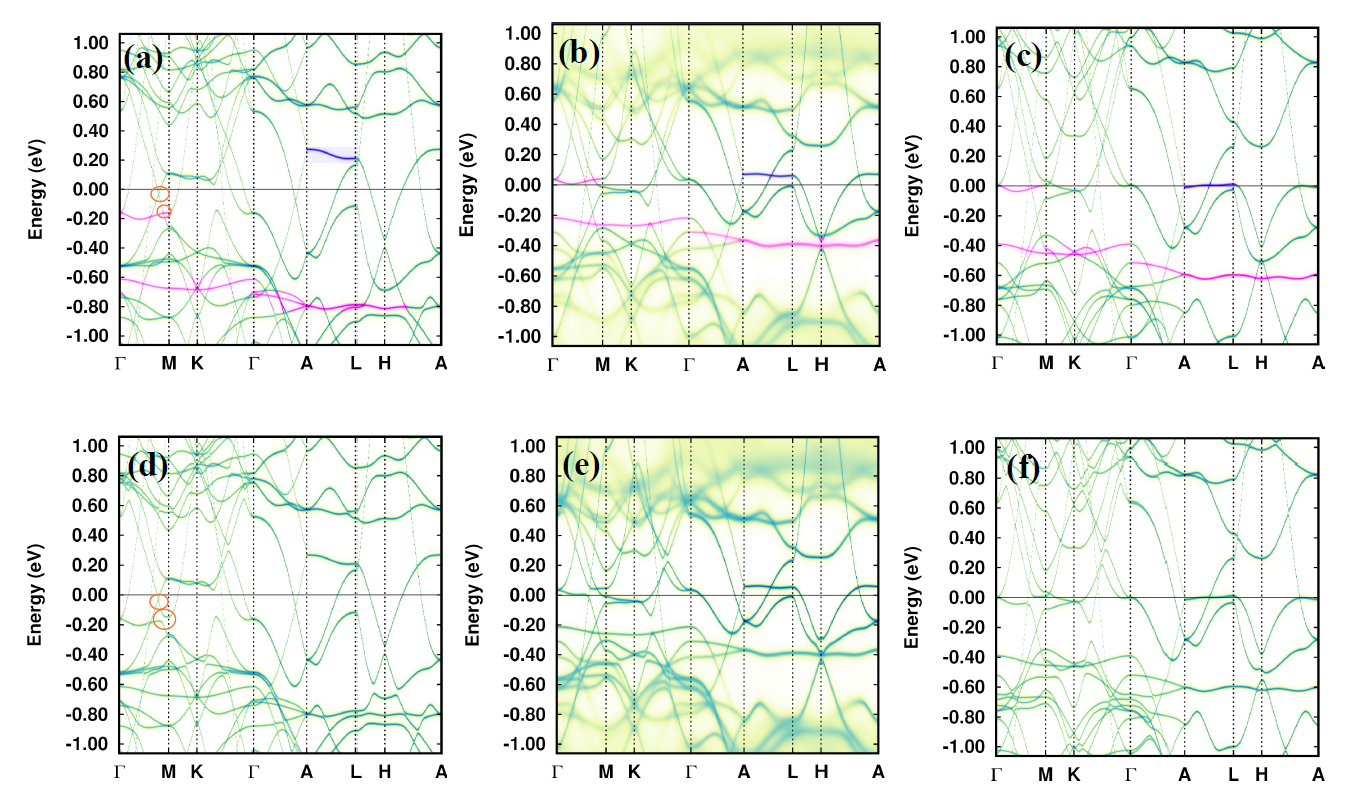}
    \caption{Spectral function of Fe$_3$Sn as obtained in DFT (a,d), DFT+DMFT (b,e) and DFT+U (c,f). Panels (a,b,c) are for calculations without SOC, while panels (d,e,f) are for calculations with SOC, for the magnetization along the easy axis (100). The magenta and blue colors are used to highlight the non-dispersive bands with majority and minority spin character, respectively. The orange color circles show the possible Weyl nodes for this specific magnetization direction.\label{fig:bands}}
\end{figure*}

\section{Details of the electronic structure}\label{sec:estructure}
The spectral functions along high-symmetry directions in the Brillouin zone are reported in Figure~\ref{fig:bands} for DFT, DFT+DMFT and DFT+U, with and without the inclusion of SOC. As in the previous Section, the data for SOC are shown for a magnetization along the easy axis (100). Focusing on the DFT results, panels (a) and (d), we see the presence of several bands with a relatively flat dispersion near the Fermi level. For convenience, we highlighted those bands in magenta and blue colors, reflecting their spin character. In agreement with the study by Belbase \textit{et al.}~\cite{PhysRevB-AHC}, we observe 2 bands spanning a larger portion of the Brillouin zone, i.e. the band at -0.6~eV along $\Gamma$-M-K-$\Gamma$ and the band at -0.8~eV along A-L-H-A. The orbital-projected spectral functions, reported in the SM~\cite{supplemental}, allow us to identify the dominant character of these bands. The local projectors are constructed by real spherical harmonics defined on a rotated reference frame $\tilde{x}$ $\tilde{y}$ $\tilde{z}$ where $\tilde{z}$ lays in the kagome plane and points toward the neighboring Sn atom, as illustrated in Figure~S3 of the SM~\cite{supplemental}. This choice is analogous to the one used for Sc$_3$Mn$_3$Al$_7$Si$_5$ in a very recent DFT+DMFT study~\cite{Mn-Kagome}. With the help of these data, we can establish that the flat band at {{-0.6~eV}} has mostly $d_{\tilde{x}\tilde{z}}$|$d_{\tilde{y}\tilde{z}}$ majority-spin character, which is consistent with recent results on the related breathing kagome material Ni$_3$In~\cite{Ye_2024-natphy}. This can be easily verified through a further rotation of the local orbitals to match the two reference frames. The other flat band, at -0.8~eV, exhibits a similar character, but has also a substantial contribution from $d_{\tilde{x}^2-\tilde{y}^2}$|$d_{\tilde{x}\tilde{y}}$ majority-spin states, in particular between L and H. Along the same high-symmetry directions, one can also observe the corresponding minority spin bands at higher energies, namely 1.6~eV and 1.4~eV, see Figure~S10 of the SM~\cite{supplemental}. Another notable flat band can be observed at -1.2~eV, mainly of $d_{\tilde{x}\tilde{z}}$|$d_{\tilde{y}\tilde{z}}$ minority-spin character. Due to the strong interlayer coupling between the two kagome layers~\cite{PhysRevLett-flatband, xu2025}, these flat bands do not extend over the entire Brillouin zone, but acquire a strong dispersion along the $\Gamma$-A direction.
In Figure~\ref{fig:bands}(a), we also observe another set of non-dispersive bands extending along smaller segments of the Brillouin zone, mainly at smaller excitation energies. The two bands highlighted in blue at about 50~meV and 200~meV come from the states determining the nodal plane between the U$^*$ points, which was suggested to drive a large anomalous Nernst effect, ten times larger than for pure Fe~\cite{chen2022large}. For these bands the hybridization between the Fe-$3d$ states and Sn-$5p$ states becomes more important, as it can be inferred from comparing the corresponding orbital-projected spectral functions in Figures~S4 and S7 of the SM~\cite{supplemental}.
In general, the presence of flat bands has been linked to the formation of ferromagnetic order in Fe stannites, especially for the analysis of Fe$_3$Sn$_2$, whose kagome features are enhanced by its quasi-two-dimensional nature~\cite{Tasaki-flatband,PhysRevLett-flatband,kinectic-ferro-kagome}. In Fe$_3$Sn, the role played by the flat bands and other non-dispersive states in the long-range order is expected to be less crucial, due to the three-dimensional magnetism emphasized above. Moreover, the analysis of Figure~\ref{fig:bands}(a) as well as the orbital projections reported in the SM~\cite{supplemental} has shown that the carriers mediating the exchange coupling belong to hybridized $3d$-$5p$ states having a marked itinerant character, at least in DFT.

The inclusion of SOC does not alter the shape and position of the non-dispersive bands significantly, but results in gap opening at the band-crossing points, which we highlighted in orange color in Figure~\ref{fig:bands}(d). This means that the Weyl nodes move away from the high-symmetry directions, as recently demonstrated by Belbase \textit{et al.}~\cite{PhysRevB-AHC}. These points were suggested to be the cause of the large anomalous Hall conductivity observed in experiment~\cite{shen2022thermodynamical}, but electronic structure calculations indicate more structured contributions across the entire Brillouin zone~\cite{PhysRevB-AHC}. By changing the magnetization axis to the hard axis (001), we observe a notable band splitting close to the Fermi level, as shown in Figures~S8 and S9 of the SM~\cite{supplemental}. This splitting stems from the dependence of the electron scattering on the spin orientation and differs from the Zeeman splitting. This phenomenon is recognized as the giant magneto band structure effect, as discussed in literature for two-dimensional CrI$_3$~\cite{GME-CrI3}. Furthermore, we observe that
the distribution of the gap openings depends significantly on the magnetization direction. This difference is connected to a different shift of the Weyl nodes under SOC, regarding both energy and location in the Brillouin zone~\cite{PhysRevB-AHC,Xuanhe-Fe3Sn}.

Next, we focus on the band structure (spectral function) obtained in DFT+DMFT, shown in panels (b) and (e) of Figure~\ref{fig:bands}. Due to the imaginary part of the self-energy, the quasiparticles acquire a finite lifetime which makes the energy bands well-defined only for energies close to the Fermi level, consistently with the Fermi liquid theory~\cite{prb-fe-Ni-DFT-DMFT}. This is also evident from the PDOS shown in Figure~\ref{fig:FePdos}, where no structure can be resolved below -2~eV and above +2~eV. The real part of the self-energy renormalizes the DFT eigenvalues, which is particularly evident in Figure~\ref{fig:bands}(b) for the non-dispersive bands highlighted in blue, which move much closer to the Fermi level. A similar renormalization has been reported recently for other kagome metals, such as Mn$_3$Sn~\cite{yu2022} and AV$_3$Sb$_5$ (A = K, Cs, Rb)~\cite{PRL-Kagome}.
The proper flat bands, highlighted in red, also move closer to the Fermi level. All these shifts are due to the combination of Fermi liquid renormalization, which induces a band narrowing~\cite{kunes2009}, and a rigid shift of the Fe-$3d$ orbitals, driven by a larger orbital polarization and a decreased spin splitting, as emphasized in the Section~\ref{sec:magnetism}. To compare these two effects, we can inspect the results obtained in DFT+U, shown in panels (c) and (f) of Figure~\ref{fig:bands}. Since these plots do not contain the effect of Fermi liquid renormalization, we can conclude that the latter is quite substantial in the correlated electronic structure obtained from DFT+DMFT, at least in reference to the energy scale we are focusing on. By extracting the quasiparticle weight from the self energy, we obtain a mass renormalization $m^*/m_{\text{e}}$ of 1.31 for majority spin and 1.25 for minority spin. These values are in line with those obtained for ferromagnetic kagome metals such as AV$_3$Sb$_5$ (A = K, Cs, Rb)~\cite{liuPRB22,PRL-Kagome,zhaoPRB21} and ScV$_6$Sn$_6$~\cite{yuPRB24}, but are slightly smaller than those expected for Mn$_3$Sn~\cite{Mn-Kagome,yu2022}. This is consistent with previous findings that the magnetic fluctuations in Mn tend to be larger than in Fe, as the former is closer to having a half-filled $3d$ shell~\cite{dimarco09prb79,dimarco09epjb72473}.
For completeness, the orbital-decomposed renormalization is reported in Table~\ref{tab:massren}.
\begin{table}[t]
\begin{tabular}{ccccc}
\hline
  \: setup \:  & spin & \: $d_{\tilde{z}^2}$ \: & \: $d_{\tilde{x}^2-\tilde{y}^2}$|$d_{\tilde{x}\tilde{y}}$ \: & \:  $d_{\tilde{x}\tilde{z}}$|$d_{\tilde{y}\tilde{z}}$ \:   \\
\hline
\: default \: & \: minority \: &  1.29 & 1.27 & 1.21 \\
\: default \: & \: majority \: &  1.25 & 1.32 & 1.34  \\
\: reduced \: & \: minority \: &  1.20 & 1.18 & 1.15  \\
\: reduced \: & \: majority \: &  1.16 & 1.19 & 1.20  \\
\hline
\end{tabular}
\caption{Mass renormalization $m^*/m_{\text{e}}$ as calculated in DFT+DMFT using the default Coulomb interaction parameters, as well as reduced values $U=2.0$~eV and $J=0.65$~eV, which were shown to provide a good quantitative agreement for the MAE.\label{tab:massren}}
\end{table}

Interestingly, for both DFT+DMFT and DFT+U, the band segments along $\Gamma$-M and M-K are predicted to cross the Fermi level, while those along A-L move at very small excitation energies. However, we should stress that this effect is very sensitive to the precise choice of the Coulomb interaction parameters. For example, the DFT+DMFT results for values of $U$ and $J$ that provide the optimal MAE, discussed in Section~\ref{sec:magnetism}, do not exhibit these changes (data not shown). This uncertainty, as well as having neglected defects of any sort, makes it impossible to provide a quantitative analysis of how far these non-dispersive bands are going to be located from the Fermi level. Nevertheless, our results highlight the following fundamental aspects: 1) a series of non-dispersive bands are located in the very vicinity of the Fermi level; 2) these non-dispersive bands are strongly affected by the local Coulomb repulsion. This suggests that one could induce a variety of Lifshitz transitions by a small modulation of external parameters, which could take the form of a small strain (to change the $3d$ bandwidth) or doping charge carriers (to change the position of the Fermi level). We believe that these aspects are behind the large sensitivity to doping and growth conditions exhibited by the magnetic properties, which have prevented the stabilization of the MAE along the out-of-plane direction~\cite{FAYYAZI2019126}. Interestingly, the formation of unexpected hole pockets and band crossing analogous to the one observed here has recently been reported by laser-based micro-focused ARPES measurements of Fe$_3$Sn$_2$~\cite{ekahana2024nature}.

Finally, we focus on the spectral functions obtained in DFT+DMFT and DFT+U when SOC is included, as observed in Figure~\ref{fig:bands}(e) and Figure~\ref{fig:bands}(f) respectively. We can see that the band crossing closest to the M point, corresponding to a Weyl point in the analysis by Belbase \textit{et al.}~\cite{PhysRevB-AHC}, moves off the high-symmetry direction, leading to the opening of a gap precisely at the Fermi level. Conversely, the band crossing close to it just moves to lower energies (-0.2~eV for both DFT+DMFT and DFT+U).
This new configuration may alter the contributions of these points to the Berry curvature, thus bringing a quantitative change to the topological properties. This may provide another route to explain the remaining discrepancy reported between calculated and theoretical values for the anomalous Hall conductivity in both single crystals~\cite{low2023,PhysRevB-AHC} and polycrystalline samples~\cite{chen2022large,Xuanhe-Fe3Sn}.
The changes induced in the interlayer coupling, visible through an additional crossing point along the $\Gamma$-A direction with respect to DFT are likely behind the drastic increase of the Dzyaloshinskii-Moriya interaction, reported in Figure~\ref{fig:jijs}(b).

\section{Conclusions}\label{sec:conclusions}
In conclusion, we conducted a comprehensive theoretical investigation of the electronic structure and magnetic properties of the three-dimensional breathing-kagome ferromagnet Fe$_3$Sn, using DFT, DFT+DMFT, and DFT+U. We calculated both intralayer and interlayer magnetic exchange interactions between the magnetic moments at the Fe sites, along with their respective band structures (spectral functions). Our results show that DFT+DMFT can be used to accurately reproduce the experimentally reported value of the MAE and to determine the most important contributions to the ferromagnetic coupling, including the presence of a 2-3\% anisotropic contribution in the form of the Dzyaloshinskii-Moriya interaction.
When the local coulomb interaction is considered explicitly, via DFT+DMFT and DFT+U, the quasiparticle spectra undergo significant changes, with a series of non-dispersive bands moving closer to the Fermi energy and even crossing it. Albeit this last conclusion depends on the particular choice of Coulomb interaction parameters, which is difficult to ascertain from first principles, these results point to a high tunability of the magnetic response, already reported for Fe$_3$Sn$_2$~\cite{PhysRevB.101.161114}, and in line with the high sensitivity to doping and structural variations~\cite{Fe3Sn-RSPt,FAYYAZI2019126}. Overall, topological Lifshitz transitions are expected to be tunable by small stimuli, such as strain and doping, which may affect the anomalous Hall and Nernst effects substantially. A key prerequisite for clearly observing this tunability is that the system does not undergo a significant phase transition, as e.g. due to emergent structural distortions or changes in the magnetic ground state.


\section*{Acknowledgments}
We would like to thank H.S. Kim, I.E. Brumboiu, L. Prodan, and V. Tsurkan for insightful discussions and technical help. Computational work was performed on resources provided by the National Academic Infrastructure for Supercomputing in Sweden (NAISS), partially funded by the Swedish Research Council through grant agreement no. 2022-06725.
We also acknowledge Polish high-performance computing infrastructure PLGrid for awarding this project access to the LUMI supercomputer, owned by the EuroHPC Joint Undertaking, hosted by CSC (Finland) and the LUMI consortium through PLL/2023/04/016450.
I.D.M. acknowledges financial support from the European Research Council (ERC), Synergy Grant FASTCORR, Project No. 854843, as well as from the STINT Mobility Grant for Internationalization (grant no. MG2022-9386).
L.C. acknowledges financial support from the Deutsche Forschungsgemeinschaft (DFG, German Research Foundation) through TRR 360, project no. 492547816.
This research is also part of the project No. 2022/45/P/ST3/04247 co-funded by the National Science Center of Poland and the European Union's Horizon 2020 research and innovation program under the Marie Skodowksa-Curie grant agreement no. 945339.\nocite{OpenDataFe3Sn}

%

\end{document}


\title{Supplemental material for the manuscript entitled \\ ``Exploring strong electronic correlations in the breathing kagome metal Fe$_3$Sn''}
\author{Shivalika Sharma}
\email{shivalika.sharma@umk.pl}
\affiliation{Institute of Physics, Nicolaus Copernicus University, 87-100 Toru\'n, Poland}
\author{Liviu Chioncel}
\affiliation{Theoretical Physics III, Center for Electronic Correlations and Magnetism, Institute of Physics, University of	Augsburg, 86135 Augsburg, Germany}
\affiliation{Augsburg Center for Innovative Technologies, University of Augsburg, 86135 Augsburg, Germany}

\author{Igor {Di Marco}}
\email{igor.dimarco@physics.uu.se}
\email{igor.dimarco@umk.pl}
\affiliation{Institute of Physics, Nicolaus Copernicus University, 87-100 Toru\'n, Poland}\affiliation{Department of Physics and Astronomy, Uppsala University, Uppsala 751 20, Sweden}
\date{\today}

\begin{abstract}
In this Supplemental Material, we include additional data on the interatomic exchange coupling, the magnetic anisotropy, and the spectral functions, including orbital-resolved projections.
\end{abstract}

{
\let\clearpage\relax
\maketitle
}
\beginsupplement
\section{Isotropic exchange coupling in absence of spin-orbit coupling}
\begin{figure*}[h!]
     \includegraphics[width=1\textwidth]{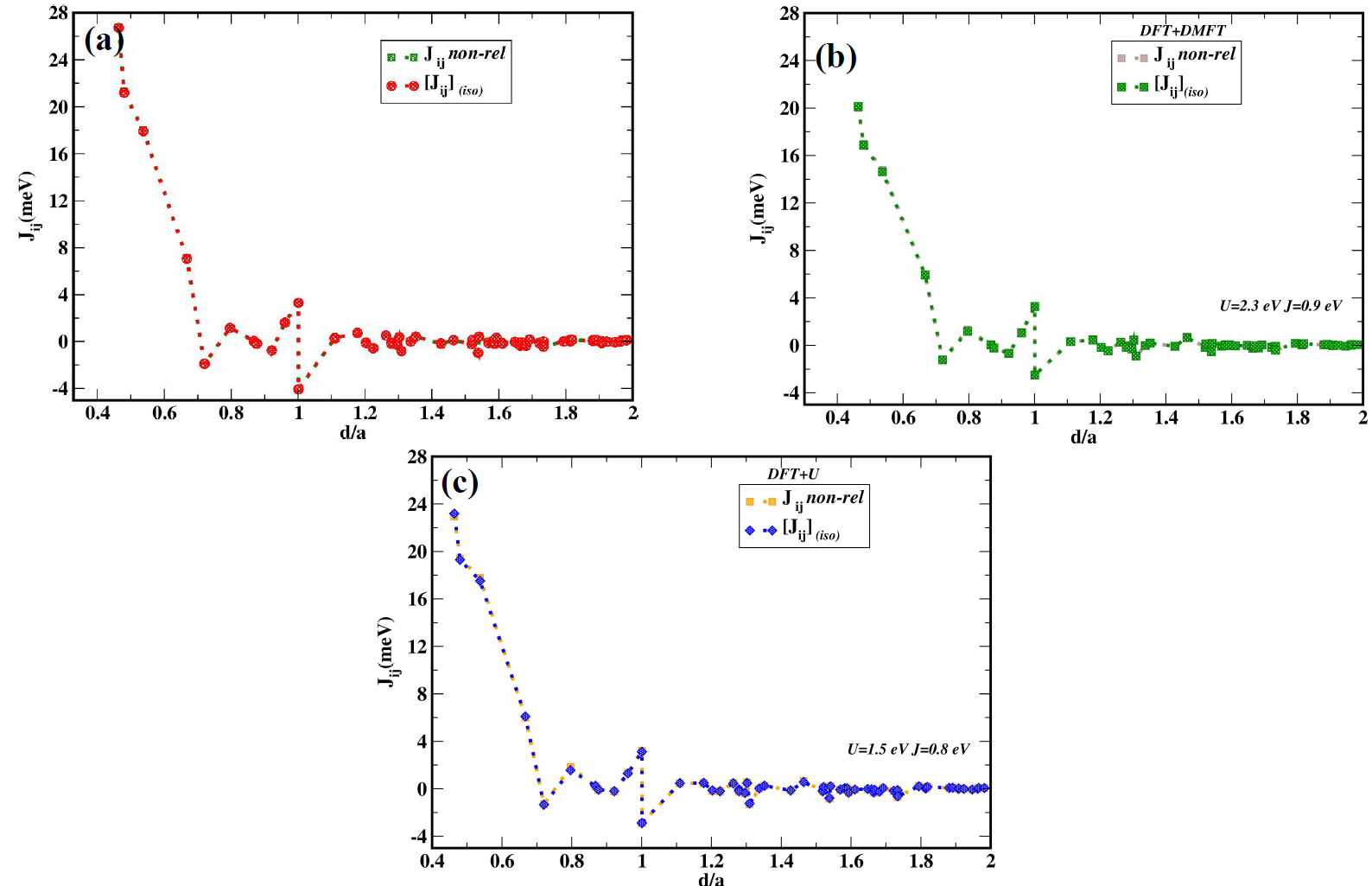}
    \caption{Comparison between the interatomic exchange parameters calculated without SOC and the isotropic part of the magnetic exchange tensor calculated with SOC, as obtained for (a) DFT, (b) DFT+DMFT, and (c) DFT+U methods.}
    \label{fig:jijcomparison}
\end{figure*}
\begin{figure*}[h!]
     \includegraphics[width=0.5\textwidth]{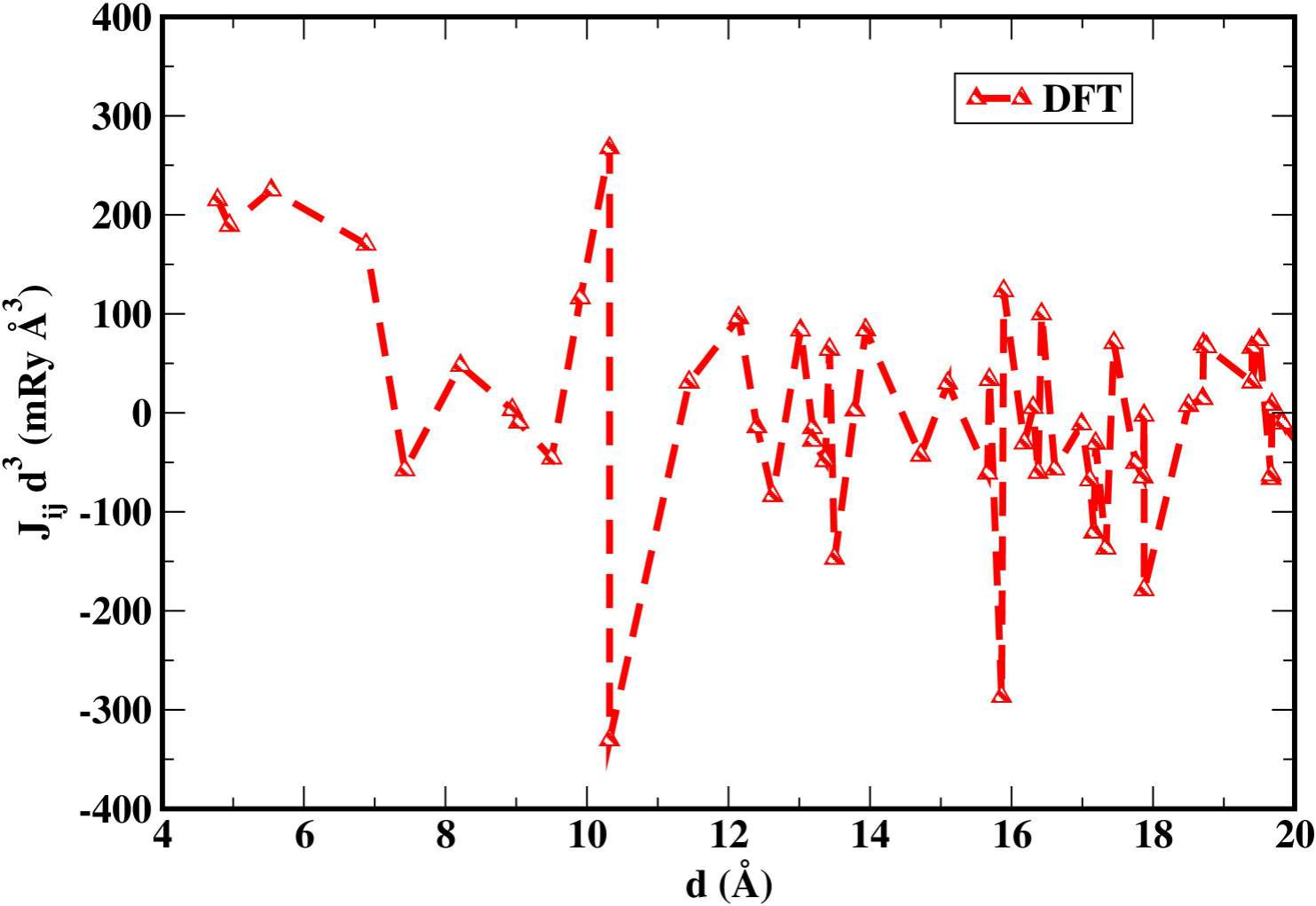}
    \caption{Interatomic exchange parameters J$_{ij}$ times the cube of the interatomic distance d$_{ij}$ as a function of d$_{ij}$. Results obtained for DFT calculations. The oscillations above 10~{\AA} seem to maintain a constant amplitude, in agreement with the expected RKKY scaling (see main text for a more detailed discussion).}
    \label{fig:Jij_longrange}
\end{figure*}
\newpage
\section{Orbital-projected spectral functions}
\begin{figure*}[h!]
\includegraphics[scale=0.55]{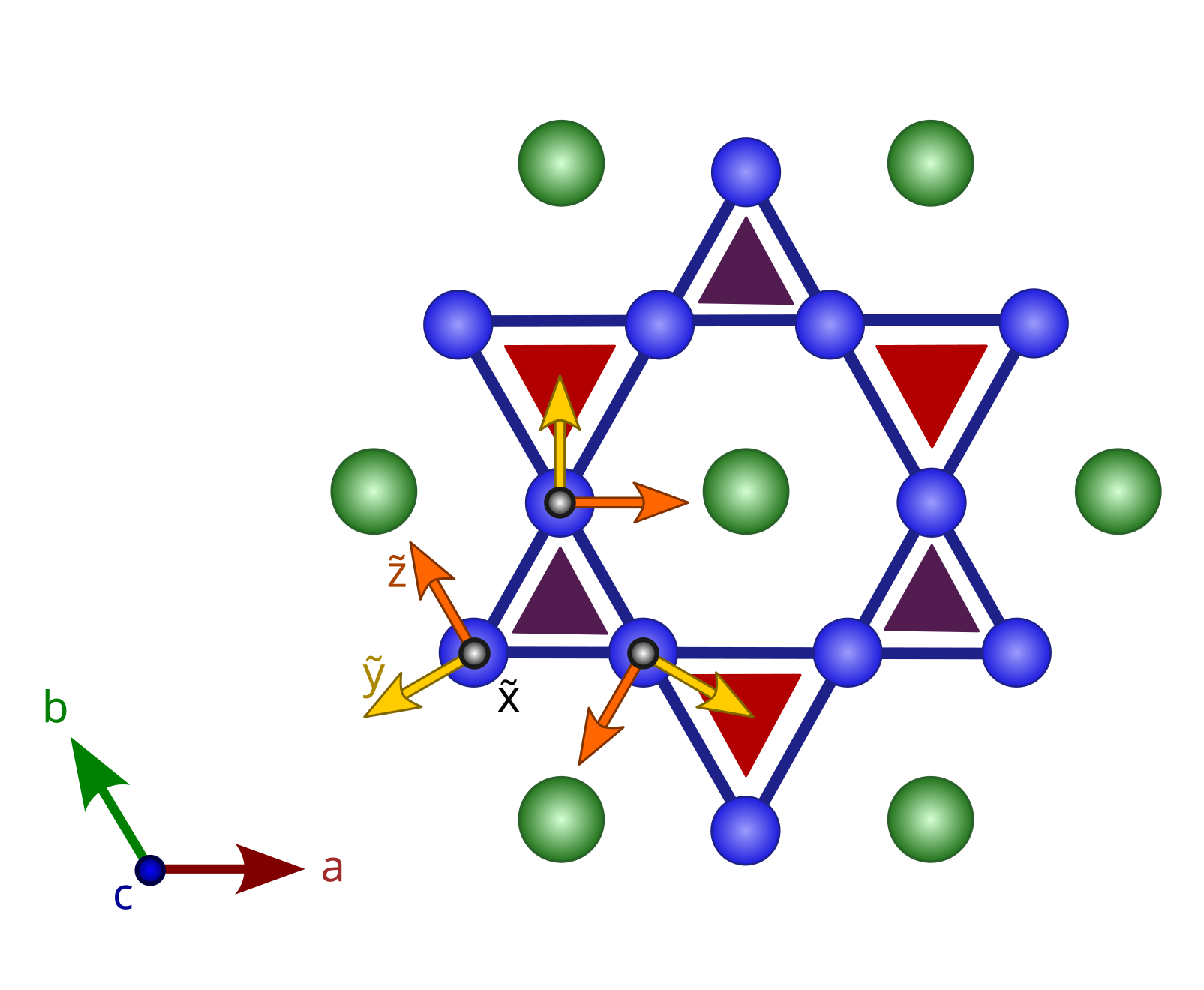}
  \caption{Local reference frame $\tilde{x}$, $\tilde{y}$, $\tilde{z}$ for the construction of the projectors onto Fe-$3d$ orbitals for the three different Fe atoms composing a single kagome layer in Fe$_3$Sn. Fe and Sn atoms are represented as blue and green spheres, respectively. The two differently sized equilateral triangles composing the breathing kagome lattice are emphasized, in dark red and violet.}
\end{figure*}
\begin{figure*}[h!]
\includegraphics[scale=0.55]{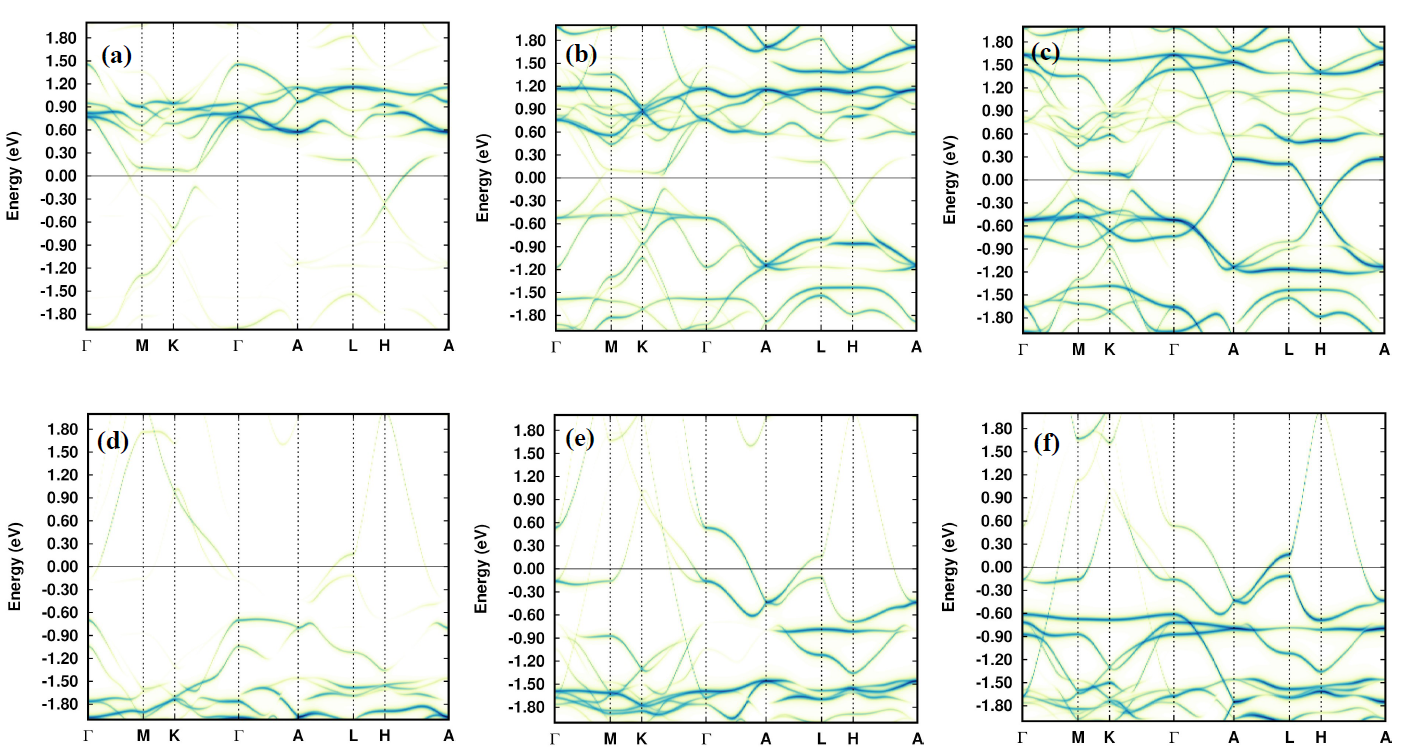}
  \caption{Orbital-projected spectral functions for the Fe-$3d$ states along high-symmetry directions in the Brillouin zone, as obtained from DFT calculations without SOC. (a)-(d) $d_{\tilde{z}^2}$; (b)-(e) average of $d_{\tilde{x}^2-\tilde{y}^2}$ and $d_{\tilde{x}\tilde{y}}$; (c)-(f) average of $d_{\tilde{x}\tilde{z}}$ and $d_{\tilde{y}\tilde{z}}$. Top and bottom panels show minority and majority spins, respectively. The Fermi level is at zero energy.}
\end{figure*}
\begin{figure*}[h!]
\includegraphics[scale=0.55]{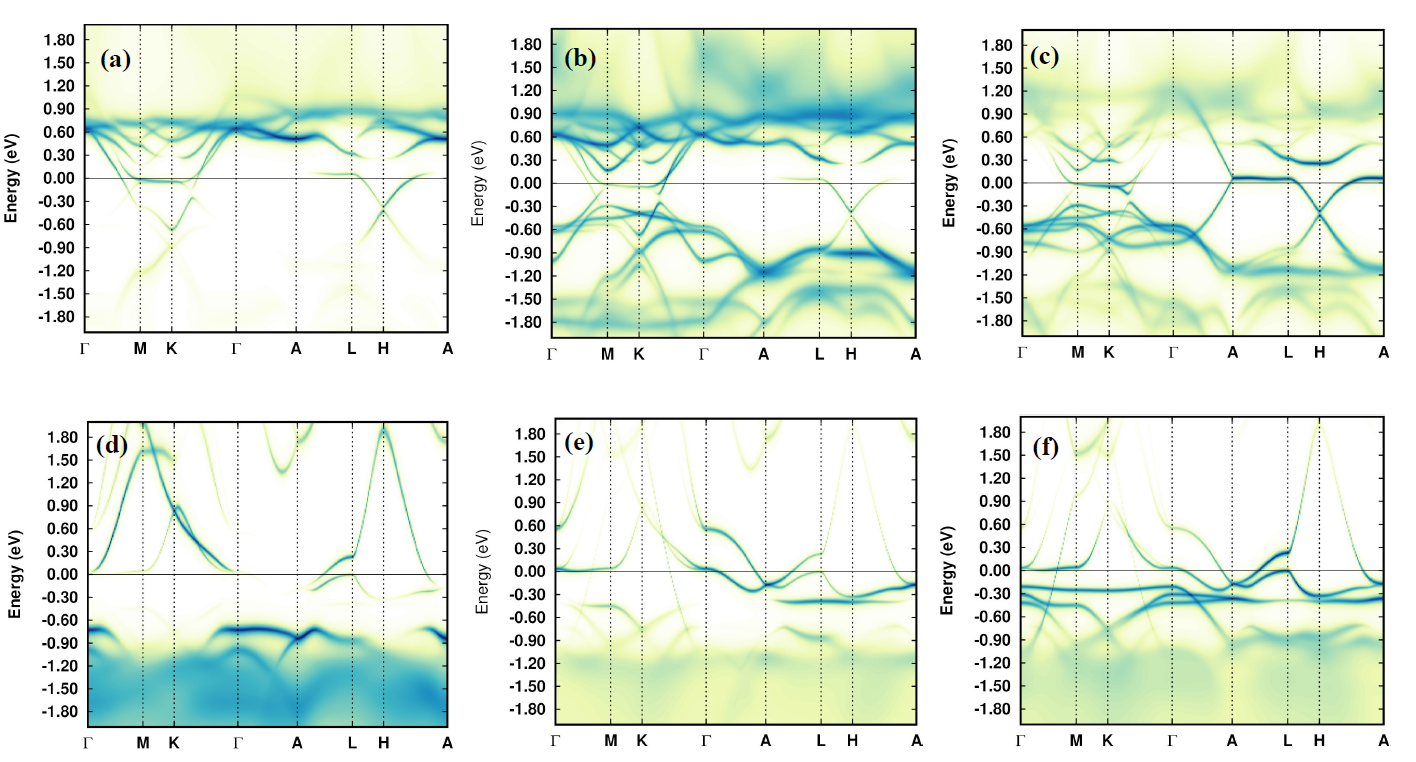}
  \caption{Orbital-projected spectral functions for the Fe-$3d$ states along high-symmetry directions in the Brillouin zone, as obtained from DFT+DMFT calculations without SOC. The Coulomb interaction parameters used for the Fe-$3d$ states are $U=2.3$~eV and $J=0.9$~eV. (a)-(d) $d_{\tilde{z}^2}$; (b)-(e) average of $d_{\tilde{x}^2-\tilde{y}^2}$ and $d_{\tilde{x}\tilde{y}}$; (c)-(f) average of $d_{\tilde{x}\tilde{z}}$ and $d_{\tilde{y}\tilde{z}}$. Top and bottom panels show minority and majority spins, respectively. The Fermi level is at zero energy.}
\end{figure*}
\begin{figure*}[h!]
\includegraphics[scale=0.55]{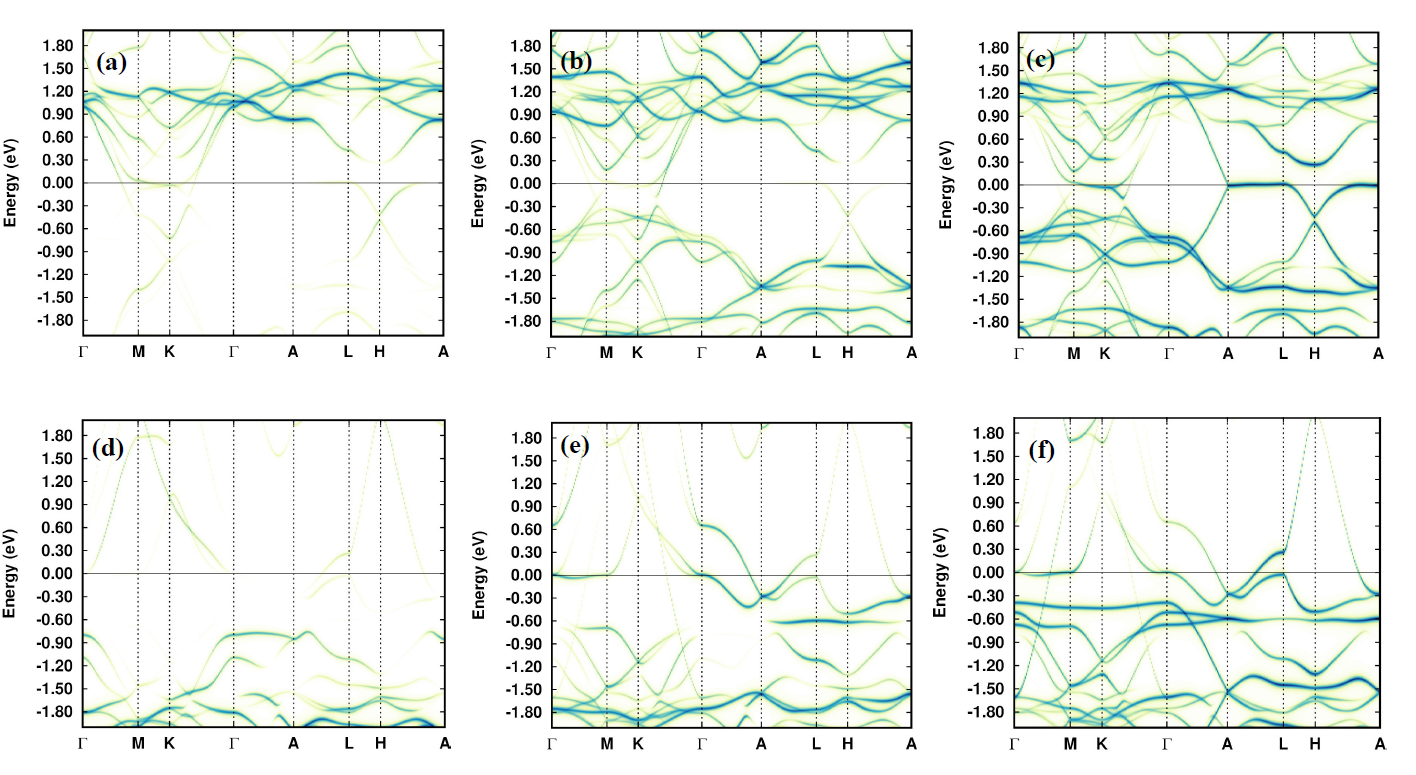}
  \caption{Orbital-projected spectral functions for the Fe-$3d$ states along high-symmetry directions in the Brillouin zone, as obtained from DFT+U calculations without SOC. The Coulomb interaction parameters used for the Fe-$3d$ states are $U=1.5$~eV and $J=0.8$~eV. (a)-(d) $d_{\tilde{z}^2}$; (b)-(e) average of $d_{\tilde{x}^2-\tilde{y}^2}$ and $d_{\tilde{x}\tilde{y}}$; (c)-(f) average of $d_{\tilde{x}\tilde{z}}$ and $d_{\tilde{y}\tilde{z}}$. Top and bottom panels show minority and majority spins, respectively. The Fermi level is at zero energy.}
\end{figure*}

\begin{figure}[h!]
\includegraphics[scale=0.55]{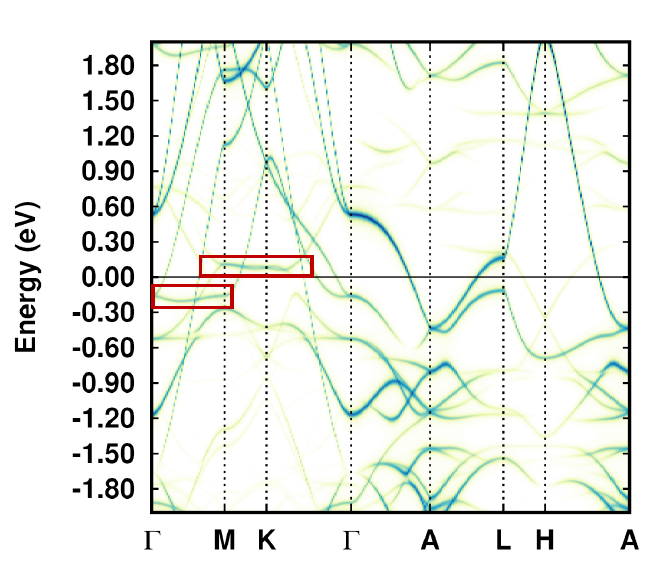}
  \caption{Orbital-projected spectral function of Sn-$5p$ states, as obtained from DFT calculations without SOC. The bands enclosed in the red rectangles are strongly hybridized with the Fe-$3d$ states. The Fermi level is at zero energy.}
\end{figure}

\newpage
\begin{figure}[h!]
\centering
\includegraphics[scale=0.55]{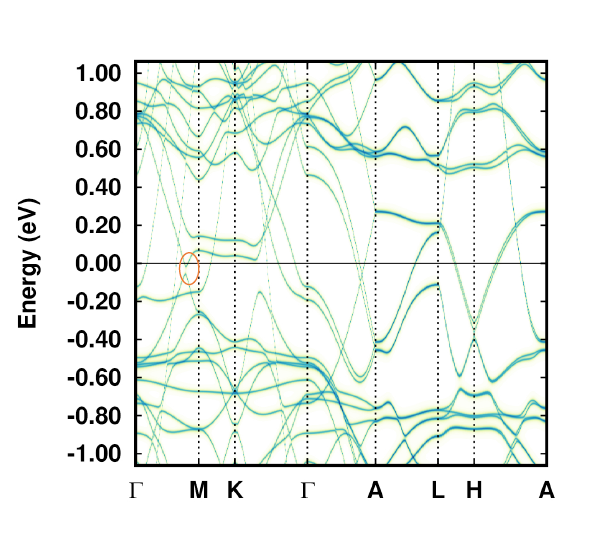}
 \caption{Spectral function (band structure) of Fe$_3$Sn as obtained in DFT with SOC, and magnetization along the (001) direction. The orange circle shows the gap forming at the Weyl node, amounting to about 0.04~eV. The Fermi level is at zero energy.}
\end{figure}
\begin{figure}[h!]
\includegraphics[scale=0.19]{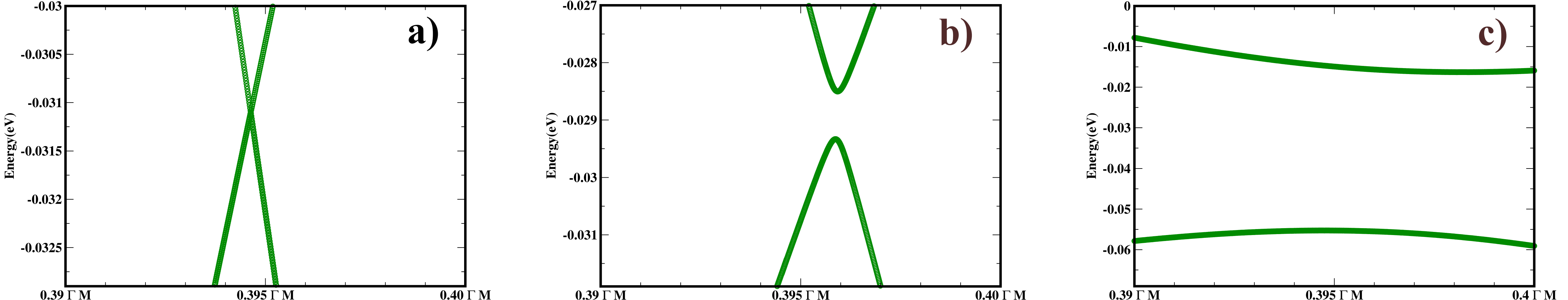}
  \caption{Magnified view of the band structure at the Weyl point for DFT calculations a) without SOC, b) with SOC and magnetization along the easy axis (100), and c) with SOC and magnetization along the hard axis (0001). The Fermi level is at zero energy.}
\end{figure}

\begin{figure}[h!]
\includegraphics[scale=0.55]{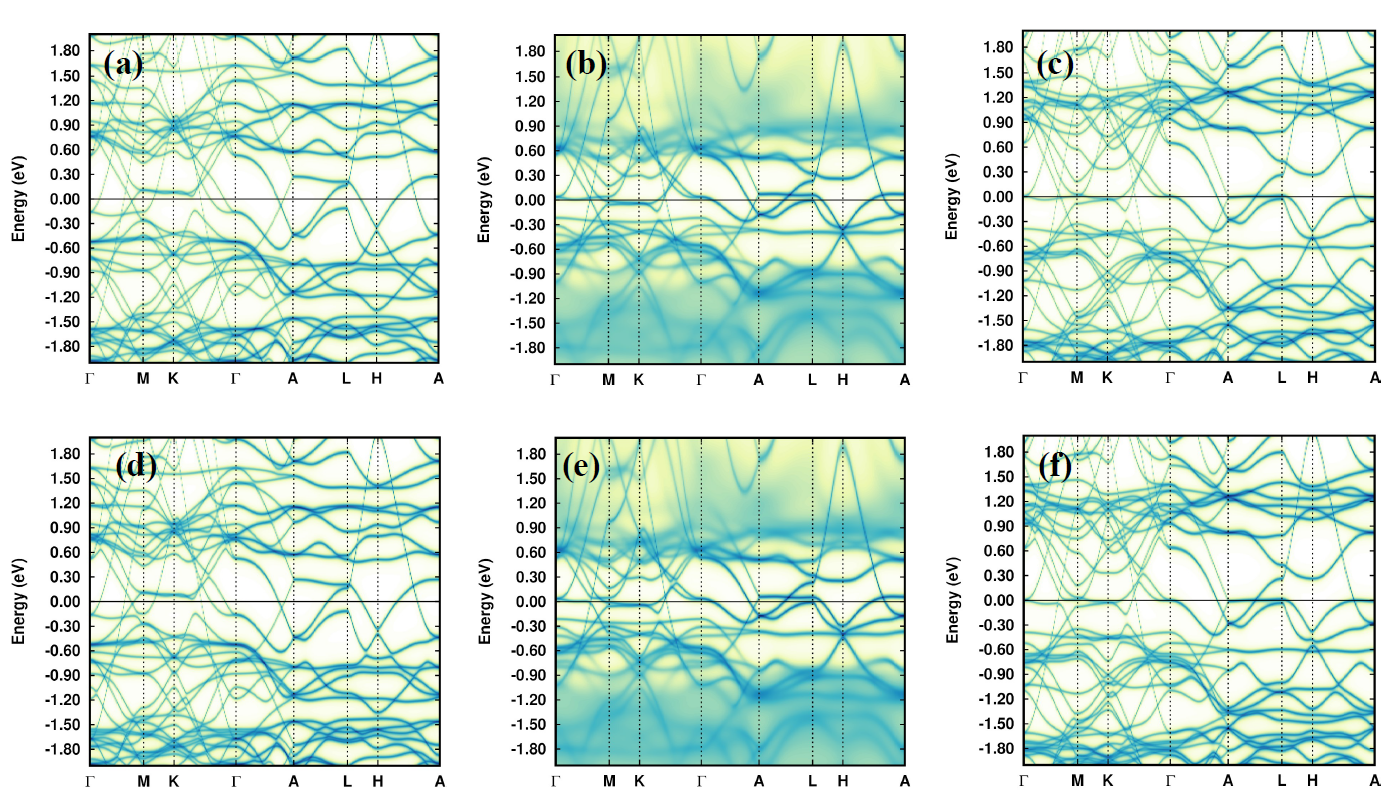}
  \caption{Spectral function of Fe$_3$Sn as obtained in DFT (a,d), DFT+DMFT (b,e) and DFT+U (c,f). Panels (a,b,c) are for calculations without SOC, while panels (d,e,f) are for calculations with SOC, for the magnetization along the easy axis (100). These plots are analogous to Figure~4 of the main manuscript, but for a larger energy range. The Fermi level is at zero energy.}
\end{figure}

\clearpage
\section{Dependence on the Coulomb interaction parameters}
\begin{figure*}[h!]
\centering
    \includegraphics[scale=0.5]{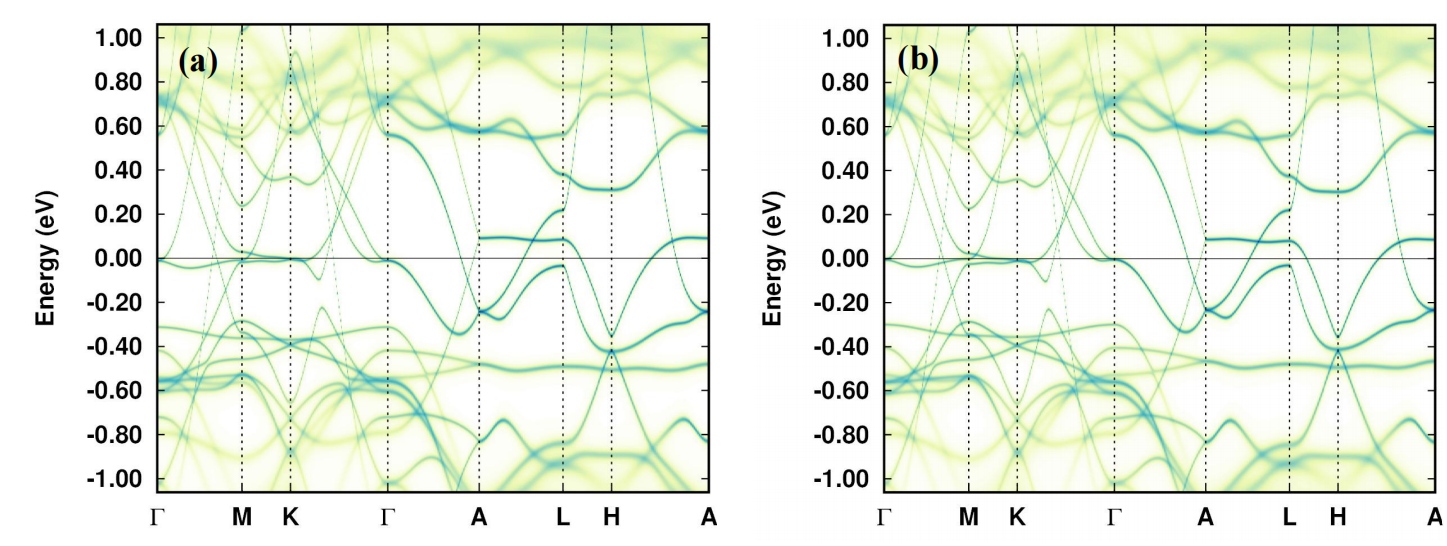}
	\caption{\label{fig:bandsujvar}Comparison of spectral functions (correlated band structure) obtained in DFT+DMFT for $U=2.0$ eV and $J=0.60$~eV (a) and  $J=0.65$~eV (b). Calculations without SOC.}
\end{figure*}

\begin{table}[t]
\caption{Relative energy of two different magnetization directions of Fe$_3$Sn in DFT+DMFT for different values of the Coulomb interaction parameters $U$ and $J$. Energies are given in meV and per formula unit (f.u.). The spin and orbital magnetic moments of the Fe-3$d$ states are also shown.}
   \centering
   \begin{tabular}{|c|c|c|c|c|}
   \hline
    \multirow{2}{*}{{$U$ (eV), $J$ (eV)}} & \multirow{2}{*}{States} &  \multirow{2}{*}{Energy (meV) / f.u.} &  \multirow{2}{*}{$\mu^s_{Fe}$} \: & \:  \multirow{2}{*}{$\mu^o_{Fe}$}  \\
    & & & &\\\hline
     \multirow{2}{*}{\text{3.00, 1.20}}&\text{FM[100]}&0&2.25 &0.10\\
    &\text{FM[001]}&3.4 &2.25&0.10\\
    \hline
    \multirow{2}{*}{\text{2.30, 0.90}}&\text{FM[100]}&0&2.31 &0.10\\
    &\text{FM[001]}&2.13 &2.31&0.10\\
    \hline
    \multirow{2}{*}{\text{1.15, 0.45}}&\text{FM[100]}&0&2.40 &0.09\\
    &\text{FM[001]}&1.89 &2.40&0.09\\
    \hline
    \multirow{2}{*}{\text{1.00, 0.40}}&\text{FM[100]}&0&2.41 &0.09\\
    &\text{FM[001]}&1.66 &2.41&0.09\\
    \hline
    \multirow{2}{*}{\text{0.00, 0.00}}&\text{FM[100]}&0&2.45 &0.07\\
    &\text{FM[001]}&0.64 &2.45&0.07\\
    \hline
    \end{tabular}
    \label{tab:ujvariations}
\end{table}
To better investigate the dependence of the magnetic anisotropy energy (MAE) on the variation of the Coulomb interaction parameters we performed additional DFT+DMFT calculations. For simplicity, we kept a fixed $U/J$ ratio, which covers previous choices of $U$ and $J$ values employed in DFT+DMFT studies of bcc Fe with perturbative solvers (see main text). The results of these calculations are shown in the Table~\ref{tab:ujvariations}. We observe that a larger Coulomb correction strengthens the stability of in-plane magnetization, while also reducing the spin magnetic moment. By keeping $U=2$ eV instead, and varying J from 0.70 eV to 0.65 eV and 0.60 eV, we calculate an MAE of 1.10, 0.45, and 0.06 meV/f.u., respectively. Interestingly, no major differences are noticeable in the comparison of the spectral functions, as e.g. in Figure~\ref{fig:bandsujvar}.
For comments on the origin of this marked dependence of the MAE on the Hund's exchange $J$, see the main text.
